\documentclass[pra,showpacs,showkeys,twocolumn,amsmath,amssymb,superscriptaddress,preprintnumbers,floatfix]{revtex4}
\usepackage{bm,amsfonts, mathtools}
\usepackage{graphicx,color, wasysym}
\usepackage{textcomp}
\usepackage{amsmath,amssymb,latexsym,epsfig}

\def\ulamek#1#2{\mbox{\normalfont$\frac{#1}{#2}$}}

\begin{document}

\makeatletter

\title{Photoluminescence decay of silicon nanocrystals and L\'{e}vy stable distributions}

\author{G.~Dattoli}
\email{dattoli@frascati.enea.it}
\affiliation{ENEA - Centro Ricerche Frascati, via E. Fermi, 45, IT 00044 Frascati (Roma), Italy\vspace{2mm}}

\author{K.~G\'{o}rska}
\email{katarzyna.gorska@ifj.edu.pl}
\affiliation{H. Niewodnicza\'{n}ski Institute of Nuclear Physics, Polish Academy of Sciences, ul.Eljasza-Radzikowskiego 152, 
PL 31342 Krak\'{o}w, Poland\vspace{2mm}}

\author{A.~Horzela}
\email{andrzej.horzela@ifj.edu.pl}
\affiliation{H. Niewodnicza\'{n}ski Institute of Nuclear Physics, Polish Academy of Sciences, ul.Eljasza-Radzikowskiego 152, 
PL 31342 Krak\'{o}w, Poland\vspace{2mm}}

\author{K.~A.~Penson}
\email{penson@lptl.jussieu.fr}
\affiliation{Laboratoire de Physique Th\'eorique de la Mati\`{e}re Condens\'{e}e,\\
Universit\'e Pierre et Marie Curie, CNRS UMR 7600\\
Tour 13 - 5i\`{e}me \'et., B.C. 121, 4 pl. Jussieu, F 75252 Paris Cedex 05, France\vspace{2mm}}


\begin{abstract}
Recent experiments have shown that photoluminescence decay of silicon nanocrystals can be described by the stretched exponential function. We show here that the associated decay probability rate is the one-sided L\'{e}vy stable distribution which describes well the experimental data. The relevance of these conclusions to the underlying stochastic processes is discussed in terms of L\'{e}vy processes.
\end{abstract}

\maketitle

\section{Introduction}

L\'{e}vy flights and the associated distributions are object of increasing interest in physics \cite{JKlafter11} and other domains, including population migration, economy and medicine, mainly regarding the metastasis spreading in cancer diseases \cite{FFioretti11}. The one-sided L\'{e}vy processes are substantially different from conventional random walk as they give rise to distributions with longer tails than the heat-type diffusion. The dynamics underlying this type of process is not fully clarified. It is however assumed that the equations ruling the evolution of distribution belong to the family of fractional derivative Fokker-Planck equations. One of the distinctive features of sub-diffusive dynamics is the appearance of relaxation functions of the stretched exponential form $\sim\exp(-x^{\alpha})$, $0<\alpha<1$. The exponent $\alpha$ determines the order of the fractional partial differential equation involved in the description of the process. In this paper we start from the experimental finding that the photoluminescence (PL) decay in nanocrystals of silicon is characterized by a stretched exponential law. Subsequently we use the recently obtained exact solutions of L\'{e}vy one-sided law to derive the analytical expression for the PL decay function. Then we deduce from its behavior the nature of the associated L\'{e}vy process. This yields an efficient tool to treat the experimental data and furnishes a consistent physical insight into the process dynamics.  

Since the first observation of PL from nanostructured porous silicon \cite{LTCanham90, VLehmann91} there has been increasing interest in the study of the optical properties of this material. While some authors explained PL phenomena as radiative recombination of charge carriers in a quantum-confined system, the others made defect luminescence responsible for the PL in the nanocrystalline Si samples (Si-NCs). The optical properties of Si-NCs are explained in various ways which encompass exciton migration between interconnected nanocrystals \cite{LPavesi96}, variation of the atomic structure of Si-NCs of different sizes \cite{OGuillois04}, tunneling of carriers from Si-NCs to traps in non-radiative process \cite{IMihalcescu96} and many others \cite{SSwada94, JLinnros99}. However, the satisfactory physical explanations underlying the phenomenological description of the photoluminescence relaxation are still absent. In this paper we will construct the effective model of PL in Si-NCs which is based on the anomalous diffusion governed by the L\'{e}vy stable distributions and the appropriate fractional Fokker-Planck equations. We will also verify our approach by comparing with the experimental results. 

In many materials with disordered structure the relaxation phenomena have been found to follow the stretched exponent or the  Kohlrausch-Williams-Watts (KWW) function \cite{RKohlrausch54, GWilliams70}
\begin{equation}\label{E1} 
\frac{n(t)}{n_{0}} = e^{-(t/\tau_{0})^{\alpha}},
\end{equation}
with $0 < \alpha < 1$ and an effective time constant $\tau_{0}$ given by $\tau_{0} = \sqrt{\pi}\langle t\rangle_{\alpha}/[2^{2/\alpha-1}\Gamma(\ulamek{1}{\alpha} + \ulamek{1}{2})]$, where $\langle t^{\mu}\rangle_{\alpha}~=~\mathcal{N} \int_{0}^{\infty} t^{\mu}e^{-(t/\tau_{0})^{\alpha}} dt$ and $\mathcal{N} =1/ [\tau_{0} \Gamma(1+\ulamek{1}{\alpha})]$ being the normalization constant of KWW and $\Gamma(z)$ the gamma function. The KWW pattern has been observed in PL phenomena in porous Si and nanocrystals Si \cite{MDovrat04, RJWalters06, GZatryb11, GZatryb10} as well as in various amorphous materials \textit{e.g.} in polymers \cite{JTBendler84}, glass like materials near the glass transition temperature \cite{DLLeslie-Pelecky94} and so on. So far the nature of the stretched exponent is not well understood. The current interpretations of the role played by the  KWW PL decay law and by the relevant index $\alpha$ range from that of a convenient phenomenological tool spoiled of any particular physical significance \cite{WGotze91}, to a deep fundamental problem to be carefully investigated \cite{JCPhilips94}. In this paper we will discuss a semi-phenomenological model useful, in our opinion, to open a new  perspective in understanding of the physical origin of the KWW dependence of PL decay in Si-NCs.

\section{Experimental results and their modelling}

Let us consider the sample consisting of Si-NCs embedded in amorphous silica. Because the energy gap in amorphous silica is bigger (9 eV \cite{L1}) than in Si-NCs (1.5 eV \cite{GZatryb11}), we can assume that PL decay occurs in Si-NCs by means of a kind of Debye relaxation. We hypothesize the existence of de-excitation channels in Si-NCs \cite{LPavesi00, LERamos05}, so that PL decay origin can be traced back to the electron transitions from the conduction band, through a de-excitation processes where each of them is ruled by Debye relaxation. The ratio of excited emitters to the number of all possible excitations is described as the weighted average of single de-excitations:
\begin{equation}\label{E2}
\frac{n(t)}{n_{0}} = \sum_{i} e^{-u_{i} \ulamek{t}{\tau_{0}}} \varPhi(u_{i}) \Delta u_{i}, 
\end{equation}
where $u_{i}$ is an ``effective variable'' describing the $i$th de-excitation channel, $\varPhi(u_{i}) \Delta u_{i}$ is a probability density of de-excitation channels of Si-NCs and $\sum_{i} \varPhi(u_{i}) \Delta u_{i} = 1$. For infinitesimally small changes of $u_{i}$, Eq. \eqref{E2} reads
\begin{equation}\label{E3}
\frac{n(t)}{n_{0}} = \int_{0}^{\infty} e^{-u \ulamek{t}{\tau_{0}}} \varPhi(u) d u. 
\end{equation}

Following the results presented in \cite{CDelerue06, FSangghaleh13}, we postulate the equality between Eqs.~\eqref{E1} and \eqref{E3}. This means that the stretched exponent is the Laplace transform of $\varPhi(u)$ and, equivalently, $\varPhi(u)$ is given by its inverse Laplace transform \cite{KAPenson10}. Mathematically, it results in the conclusion that $\varPhi(u)$ is uniquely determined \cite{HBergstrom52} and it is given by the one-sided L\'{e}vy stable distribution (LSD) \cite{KAPenson10}. According to the common convention LSD is defined as the inverse Laplace transform \cite{HPollard46}. Namely, for $u \geq 0$ we get  \cite{KAPenson10, HPollard46}
\begin{align}\label{E6}
\varPhi_{\alpha}(u) &= \frac{1}{\pi} {\rm Im}\left\{\int_{0}^{\infty} e^{-u\xi} e^{-\xi^{\alpha} e^{-i\pi\alpha}} d\xi\right\} \\ \label{E6a}
& = \frac{1}{\pi} {\rm Im}\left\{\sum_{k=0}^{\infty} \frac{(-1)^{k}}{k!} e^{-i\pi\alpha k} \frac{\Gamma(1+k\alpha)}{u^{1+k\alpha}}\right\}
\end{align}
and $\varPhi_{\alpha}(u) = 0$ for $u< 0$. 

\begin{figure}[!h]
\begin{center}
\includegraphics[scale=0.40]{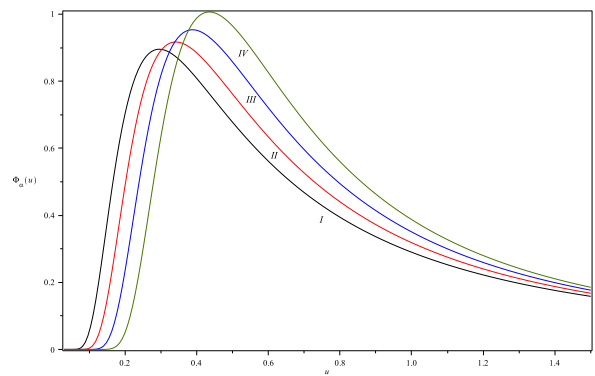}
\caption{\label{fig0} (Color online) Comparison of $\varPhi_{\alpha}(u)$ as given in Eq. \eqref{Ea} for $\alpha = 3/5$ (\textit{I}, black line), $\alpha = 19/30$ (\textit{II}, red line), $\alpha = 2/3$ (\textit{III}, blue line), and $\alpha = 7/10$ (\textit{IV}, green line).}
\end{center}
\end{figure}

The identification of the probability density $\varPhi_{\alpha}(u)$ in terms of L\'{e}vy functions is an important hypothesis. It does allow: (a) the use of well established methods to solve the integral equation in Eq. \eqref{E3} to obtain the explicit form of the function $\varPhi_{\alpha}(u)$; (b) the possible identification of the process underlying the PL decay in terms of L\'{e}vy random flights. Up-to-now Eq. \eqref{E6} was usually solved numerically (\textit{e.g.} via Stehfest algorithm). Recently, two of us found in \cite{KAPenson10, KGorska11} the exact and explicit form of LSD for rational $\alpha = l/k$, $0 < \alpha = l/k < 1$, for integers $l$ and $k$. Below, we quote the analytically derived Eq. (3) in \cite{KAPenson10}, 
\begin{equation}\label{Ea}
\varPhi_{l/k}(u) = \sum_{j=1}^{k-1} \frac{b_{j}(k, l)}{u^{1+j\ulamek{l}{k}}} {\,_{l+1}F_{k}}\left({1, \Delta(l, 1 + j\ulamek{l}{k}) \atop \Delta(k, j+1)}\Big\vert z\right),
\end{equation}
where $z=(-1)^{k-l} l^{l}/(k^{k} u^{l})$. The coefficients $b_{j}(k, l)$ are given by the ratios of Euler's gamma functions, see Eq. (4) in \cite{KAPenson10}. ${_{l+1}F_{k}}$ are the generalized hypergeometric functions whose the first (``upper'') list of parameter is equal to $1, \Delta(l, 1+ jl/k)$ and the second (``lower'') list is given via $\Delta(l, j+1)$, where $\Delta(n, a)~=~\ulamek{a}{n}, \ulamek{a+1}{n}, \ldots, \ulamek{a+n-1}{n}$. We remark that the generalized hypergeometric functions are very well implemented in the standard computer algebra systems. For reader's convenience we give in \cite{CAS} the Maple$^{\text{\textregistered}}$ syntax for $\varPhi_{l/k}(u)$ and we illustrate Eq. \eqref{Ea} for $\ulamek{l}{k}=\ulamek{3}{5}, \ulamek{19}{30}, \ulamek{2}{3}$ and $\ulamek{7}{10}$ in Fig.~\ref{fig0}.
The exact and explicit form of LSD enabled us to confirm their the following properties: existence of essential singularities at $u=0$; heavy-tailed asymptotic behavior for large $u$; all fractional moments $M_{\alpha}(\lambda)~=~\int_{0}^{\infty} u^{\lambda} \varPhi_{\alpha}(u) du$ being finite for $-\infty < \lambda < \alpha$, while the moments $\lambda \geq \alpha$, including the mean value and the variance, being infinite. Moreover, $\varPhi_{\alpha}(u)$ are infinitely divisible, closed under two types of convolution (see Eq. \eqref{E9} below for the first type and Eq. (30) in \cite{KGorska12a} for another type), and unimodal.

Note that the convolution property given in Eq. (30) in \cite{KGorska12a}, \textit{i.e.} for $0 < \gamma, \beta < 1$
\begin{equation}\label{Eb}
\varPhi_{\gamma\beta}(u) = \int_{0}^{\infty} \frac{1}{v^{1/\gamma}} \varPhi_{\gamma}\left(\frac{u}{v^{1/\gamma}}\right) \varPhi_{\beta}(v) dv, 
\end{equation}
implies that PL effect for the whole sample is equal to the weighted average of stretched exponents of single nanocrystals. More precisely, we can write
\begin{equation}\label{Ec}
\frac{n(t)}{n_{0}}  = \sum_{i} e^{-z_{i} \big(\ulamek{t}{\tau_{i}}\big)^{\gamma}} \varPsi(z_{i}) \Delta z_{i},
\end{equation}
where for $0 < \gamma < 1$, $z_{i}$ is an ``effective variable'' describing the $i$th de-excitation channel in single Si-NC, $\tau_{i}$ is an effective time constant of given channel, and $\varPsi(z) \Delta z$ is a probability density of de-excitation channels of single Si-NC. Eq. \eqref{Ec} rewritten for infinitesimally small $z_{i}$ has the form
\begin{equation}\label{Ed}
\frac{n(t)}{n_{0}} = \int_{0}^{\infty} e^{-z\big(\ulamek{t}{\tau}\big)^{\gamma}} \varPsi(z) dz.
\end{equation}
Using once again the relation between the stretched exponent and LSD we get
\begin{align}\label{Ee}
\frac{n(t)}{n_{0}} = \int_{0}^{\infty} \left[\int_{0}^{\infty}e^{-s\ulamek{t}{\tau}} g_{\gamma}\left(\frac{s}{z^{1/\gamma}}\right) \frac{ds}{z^{1/\gamma}}\right]\varPsi(z) dz, 
\end{align}
where $g_{\gamma}(x)$ is another LSD function given in Eqs.~\eqref{E6} and \eqref{Ea}. After changing the order of integration in Eq.~\eqref{Ee}, assuming that $\varPsi(z)~\equiv~\varPsi_{\beta}(z)$, $0 < \beta < 1$, is LSD and $\alpha = \gamma\beta$, and applying Eq. \eqref{Eb}, we get Eq. \eqref{E3}. That allow us to conclude that the above two ways of description the nature of stretched exponent give the same results.

\begin{figure}[!h]
\begin{center}
\includegraphics[scale=0.40]{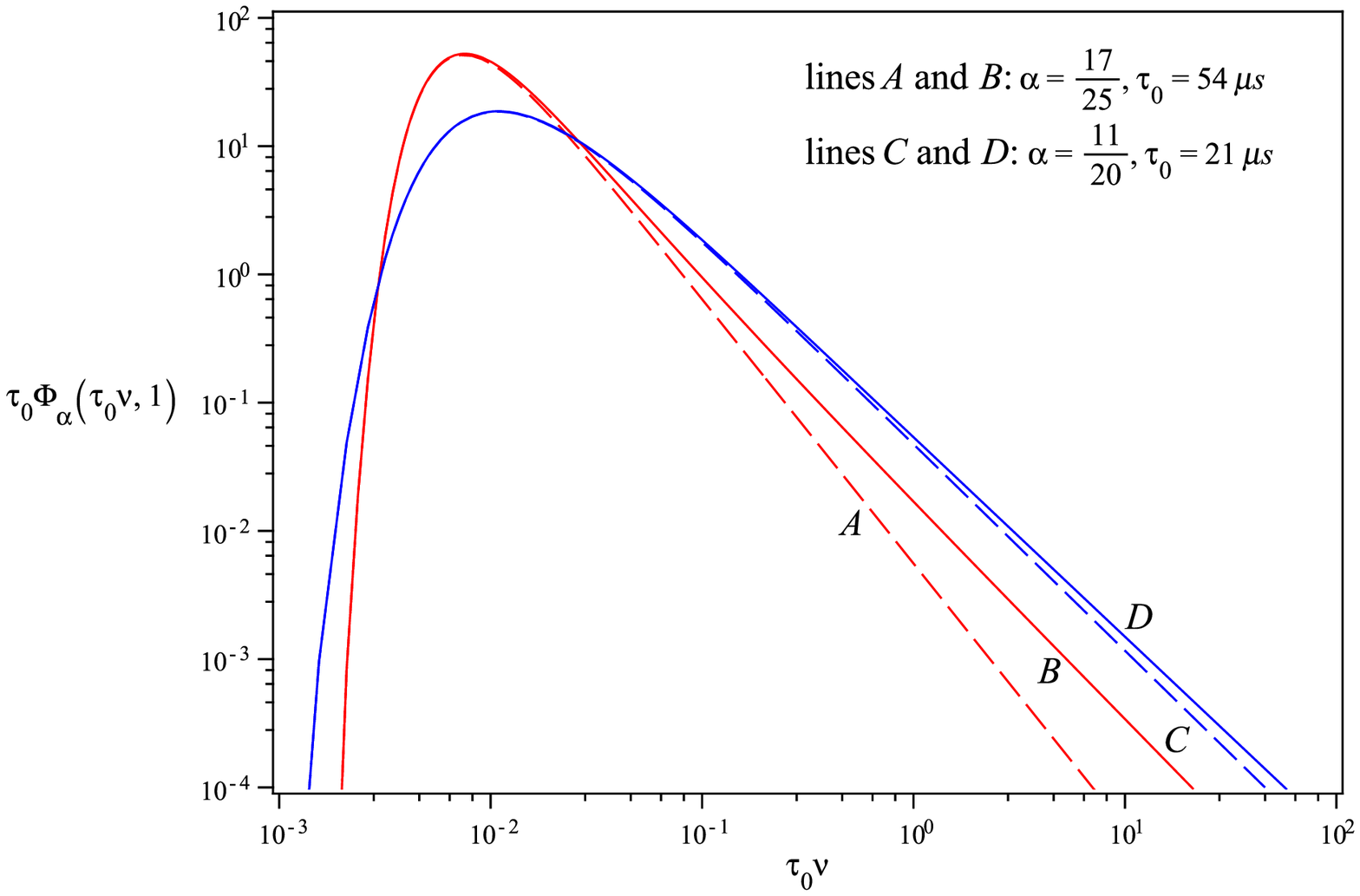}

\includegraphics[scale=0.40]{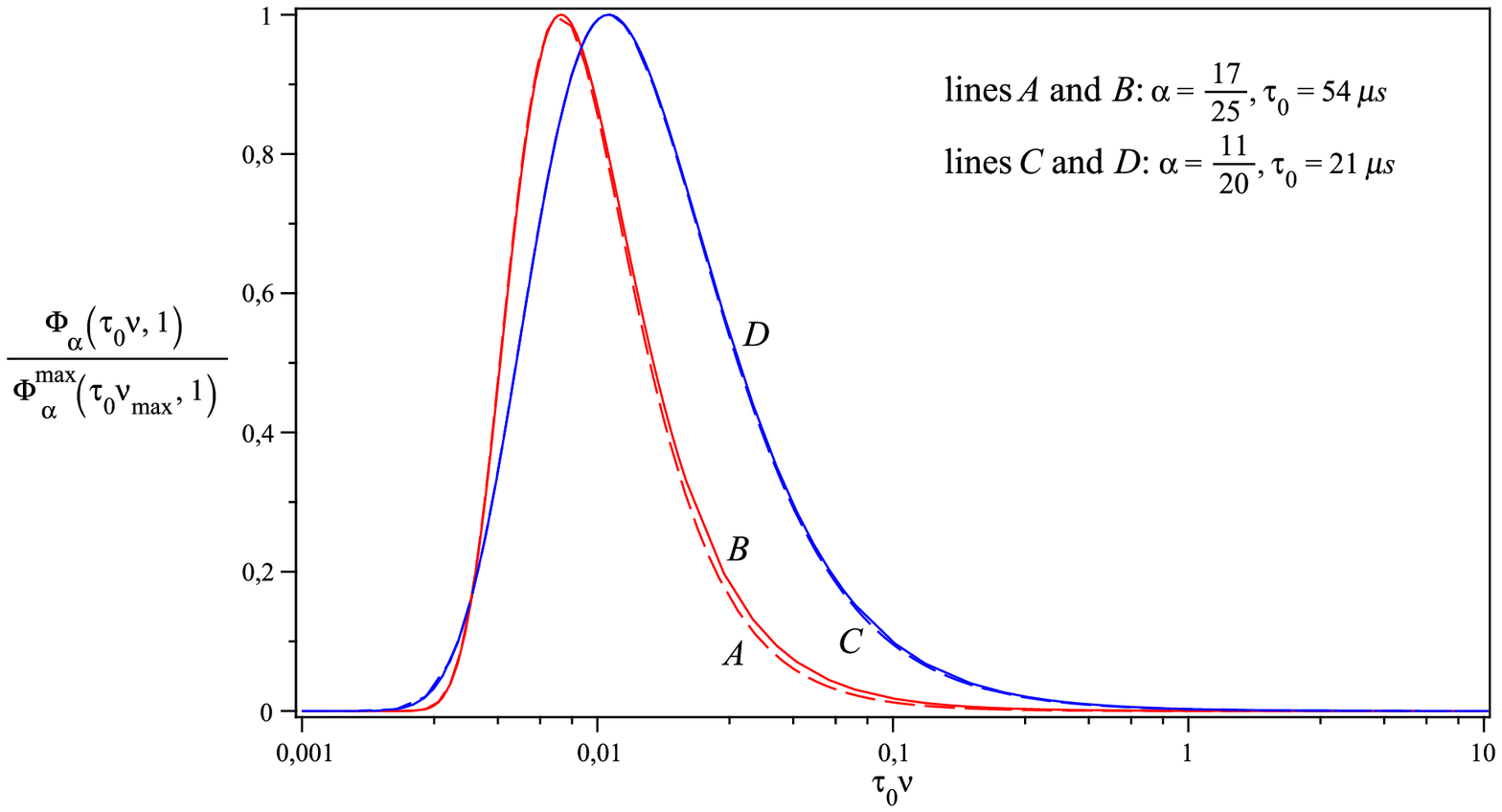}
\caption{\label{fig1} (Color online) Comparison of exact and explicit forms of LSD from Eqs. \eqref{Ea} with coefficients $b_{j}(k, l)$ given in Eq. (4) in \cite{KAPenson10} (solid lines) with the numerical fit of experimental data from Eq. (9) in \cite{GZatryb11} (dashed lines). In a) there is a double logarithmic plot and in b) there is a logarithmic plot. The values of $\alpha$ and $\tau_{0}$ pertaining to appropriate pairs of curves are indicated in the insets.}
\end{center}
\end{figure}

A benchmark of our interpretation of the PL decay probability density function comes from the experimental results of Ref. \cite{GZatryb11}. The authors used the magneton sputtering method with an hydrogen and argon mixture described by the hydrogen rate $r_{H}$ and obtained samples in which Si-NCs are embedded in amorphous SiO$_{2}$. Next, they measured PL decays under different physical conditions and got the characteristic coefficients of the stretched exponential, \textit{i.e.} $\alpha$ and $\tau_{0}$ from the numerical analysis of experimental data. As they pointed out the quantity effectively measured was not the relaxation function $\varPhi_{\alpha}(u)$ itself but rather the number of photons emitted per unit time and related to the relaxation function by the Weibull distribution \cite{WWeibull51}
\begin{align}\label{E8}
I_{\rm PL}(t) &= -\frac{1}{n_{0}} \frac{d n(t)}{d t} = \frac{\alpha}{\tau_{0}} \left(\frac{t}{\tau_{0}}\right)^{\alpha-1} e^{-(t/\tau_0)^{\alpha}} \nonumber \\[0.4\baselineskip]
& = \frac{1}{\tau_{0}} \int_{0}^{\infty} u\, e^{- u \ulamek{t}{\tau_{0}}}\, \varPhi_{\alpha}(u)\, du.
\end{align}
We remark that the median of Eq. \eqref{E8} can lead to the alternative interpretation of $\tau_{0}$, namely $\tau_{0}~=~T^{\alpha}_{1/2}/[\ln(2)]^{1/\alpha}$. The symbol $T^{\alpha}_{1/2}$ denotes the median of $I_{\rm PL}$ and it can be interpreted as the half-time of relaxation process. In \cite{GZatryb11} it has been found that $I_{PL}(t)$ depends on various parameters: the temperature, the excitation lamp wavelength $\lambda_{\rm EXC}$, the emission wavelength $\lambda_{\rm EM}$, hydrogen rate $r_{\rm H}$ and so on (for further comments see \cite{GZatryb11}). Keeping temperature, $\lambda_{\rm EXC}$ and $\lambda_{\rm EM}$ fixed and varying $r_{\rm H} = 10\%$, $30\%$ and $50\%$ the following values of $\alpha$ and $\tau_{0}$ were obtained: $\alpha\approx 0.68$ and $\tau_{0} \approx 54\mu$s, $\alpha\approx 0.57$ and $\tau_{0} \approx 30\mu$s, and $\alpha\approx 0.55$ and $\tau_{0} \approx 21\mu$s, respectively. We have ``fine tuned'' the exact solution \cite{KAPenson10} by conveniently choosing the integer ratios that reproduce the above $\alpha$ values. For readability of the Figs. \ref{fig1}a and \ref{fig1}b we compare there LSD only for two values of $\alpha$, namely $\alpha=17/25$ and $\alpha = 11/20$. The dashed lines are obtained from numerical fit of experimental data, see Eq. (9) in \cite{GZatryb11}, while the solid lines are the exact and explicit forms of LSD presented in Eqs. (3) and (4) in \cite{KAPenson10}. We see that the solid curves are consistent with the experimental results presented in \cite{GZatryb11}. Indeed, Eq. (9) in \cite{GZatryb11} is the asymptotic formula of LSD (exactly calculated from Eq. \eqref{E6}) for small values of argument, as found in Eq. (4) of \cite{JMikusinski59}. The good agreement between exact solution of Eq. \eqref{E6} and the numerical analysis of experimental results, obtained in asymptotic regime, suggests that the interface states can be described by the random energy model introduced in \cite{CDeDominicis85}, where the KWW law also appears. Denoting $\Phi^{\rm exp}_{\alpha}(y)$ and $\varPhi^{\rm asym}_{\alpha}(y; 1)$ as the distributions defined in Eq. (9) of \cite{GZatryb11} and Eq. (4) of \cite{JMikusinski59}, respectively, we can write $\Phi^{\rm exp}_{\alpha}(u) = \varPhi^{\rm asym}_{\alpha}(u)$. However, $\Phi^{\rm exp}_{\alpha}(u)$ is not normalized, \textit{i.e.} $\int_{0}^{\infty} \Phi^{\rm exp}_{\alpha}(u) du = \ulamek{1}{\sqrt{2\alpha}}$. This may account for the differences between the solid and dashed curves visible in Figs. \ref{fig1}a and \ref{fig1}b for large value of $u$. 

\section{Theoretical description}

Let us now look closer into the nature of KWW law occurring in PL in the sample where the single Si-NCs are placed in a disordered way. We propose the sub-diffusion model of an electron behavior. The electron is trapped in a single Si-NC and after some time, related to $u$, it jumps to another single Si-NC. The diffusion lengths, denoted by the new variable $\kappa$, depend on a nature of the sample, \textit{e.g.} concentration of Si-NCs in the sample related to $r_{\rm H}$. Let us assume that the variable $u$ represents the set of evolutive variables which enter into PL. By the evolutive variable we mean a quantity which controls the dynamics of the process, \textit{e.g.} continuous increase of temperature or changes of the  hydrogen rate. From the mathematical point of view such a variable plays the role of the time in the ordinary or higher-order heat equations \cite{KGorska13a}. We start with the observation that LSD satisfies the following convolution integral property
\begin{equation}\label{E9}
\int_{0}^{u} \varPhi_{\alpha}(v, \kappa) \varPhi_{\alpha}(u-v, \Delta \kappa) dv = \varPhi_{\alpha}(u, \kappa + \Delta\kappa),
\end{equation}
where
\begin{equation}\label{E7}
\varPhi_{\alpha}(u, \kappa) = \ulamek{1}{\kappa^{1/\alpha}} \varPhi_{\alpha}\big(\!\ulamek{u}{\kappa^{1/\alpha}}, 1\big) \equiv \ulamek{1}{\kappa^{1/\alpha}} \varPhi\big(\!\ulamek{u}{\kappa^{1/\alpha}}\big).
\end{equation}
Eq. \eqref{E9} can be proved after substituting Eq. \eqref{E6} into Eqs. \eqref{E7} and \eqref{E9}, changing the order of integrations, and using properties of Dirac delta. Eq. \eqref{E9} is the evolution-like integral equation of LSD which for small positive $\Delta\kappa$ can be rewritten in the differential form \cite{KGorska14}. To derive that form we study the $\Delta\kappa\to 0$ asymptotic behavior of Eq. \eqref{E9}. The first two terms of the Taylor series of r.h.s. of Eq. \eqref{E9} for $\Delta\kappa\ll 1$ read
\begin{equation}\label{E10}
\varPhi_{\alpha}(u, \kappa+\Delta\kappa) \simeq \varPhi_{\alpha}(u, \kappa) + \Delta\kappa\, \partial_{\kappa}\varPhi_{\alpha}(u, \kappa).
\end{equation}
The l.h.s of Eq. \eqref{E9} can be estimated by expressing $g_{\alpha}(u-v, \Delta\kappa)$ in terms of Eqs. \eqref{E7} and \eqref{E6a}. That gives
\begin{equation}\label{11.04.14-1}
\varPhi_{\alpha}(u, \kappa)\! \simeq\! \frac{1}{\pi}{\rm Im}\left\{\!\frac{1}{u-v}\! -\! \Delta\kappa \frac{e^{-i\pi\alpha}\Gamma(1+\alpha)}{(u-v)^{1+\alpha}}\!\right\},
\end{equation}
which after inserting it into l.h.s. of Eq. \eqref{E9} leads to
\begin{align}\label{11.04.14-2}
&{\rm l.h.s.\,\, of\,\, Eq. \eqref{E9}} = \frac{1}{\pi}{\rm Im}\left\{\int_{0}^{u}\! \frac{\varPhi_{\alpha}(v, \kappa)}{u-v} dv\right\} -\frac{\Delta\kappa}{\pi}\Gamma(\alpha) \nonumber\\
&\,\, \times {\rm Im}\left\{e^{-i\pi\alpha}\int_{0}^{u}\!\! \varPhi_{\alpha}(v, \kappa) \frac{d}{dv}(u-v)^{-\alpha} dv \right\}.
\end{align}
Applying the Sokhotski-Plemelj theorem \cite{VSVladimirov71} for the first integral in Eq. \eqref{11.04.14-2}, integrating by parts the second integral in Eq. \eqref{11.04.14-2}, and using the fact that LSD vanishes at zero, we obtain
\begin{align}\label{11.04.14-3}
&{\rm l.h.s.\,\, of\,\, Eq. \eqref{E9}} = \varPhi_{\alpha}(u, \kappa) - \frac{\Delta\kappa}{\pi}\, \Gamma(\alpha) \nonumber \\
&\times\! {\rm Im}\left\{\!\lim_{v\to u} \frac{\varPhi_{\alpha}(v, \kappa)}{(v-u)^{\alpha}} - e^{-i\pi\alpha}\!\!\! \int_{0}^{u} \frac{\varPhi'_{\alpha}(v, \kappa)}{(u-v)^{\alpha}} dv\!\right\} \\
&\,\, = \varPhi_{\alpha}(u, \kappa) - \frac{\Delta\kappa}{\Gamma(1-\alpha)} \int_{0}^{u} \frac{\varPhi'_{\alpha}(v, \kappa)}{(u-v)^{\alpha}} dv \nonumber \\[0.3\baselineskip] \label{11.04.14-4}
&\,\,- \frac{\Delta\kappa}{\pi}\, \Gamma(\alpha)\,{\rm Im}\left\{\underset{v = u}{{\rm Res}}\,\frac{\varPhi_{\alpha}(v, \kappa)}{(v-u)^{1+\alpha}}\right\},
\end{align}
where $\varPhi'_{\alpha}(v, \kappa) = \ulamek{d}{d v} \varPhi_{\alpha}(v, \kappa)$. In obtaining Eq. \eqref{11.04.14-4} the first Euler's reflection formula $\Gamma(z) \Gamma(1-z) = \pi/\sin(z\pi)$ has been used. The symbol $\underset{v = u}{{\rm Res}} f(v)$ denotes the residue of a function $f(v)$ along a path enclosing its singularities at $v=u$. Using Eq. \eqref{E6a} it is easy to show that the residue of $\varPhi_{\alpha}(v, \kappa)/(v-u)$ at $v=u$ is a real function. That leads to vanishing of the last term in Eq. \eqref{11.04.14-4}. Thus, l.h.s. of Eq. \eqref{E9} reads
\begin{equation}\label{11.04.14-5}
{\rm l.h.s.\,\, of\,\, Eq. \eqref{E9}} = \varPhi_{\alpha}(u, \kappa) - \Delta\kappa\, \partial^{\alpha}_{u} \varPhi_{\alpha}(u, \kappa),
\end{equation}
where $\partial_{x}^{\alpha} f(x)=\Gamma(1-\alpha)^{-1} \int_{0}^{x} f'(y)(x-y)^{-\alpha} dy$ is the Caputo fractional derivative \cite{IPodlubny99}. Consequently, comparing Eq. \eqref{E10} with Eq. \eqref{11.04.14-5}, the integral evolution equation \eqref{E9} can be rewritten as
\begin{equation}\label{E16}
\partial_{\kappa}\varPhi_{\alpha}(u, \kappa) = - \partial_{u}^{\alpha} \varPhi_{\alpha}(u, \kappa).
\end{equation}
It is the anomalous diffusion equation \cite{VVUchaikin00}.

\begin{figure}[!h]
\begin{center}
\includegraphics[scale=0.40]{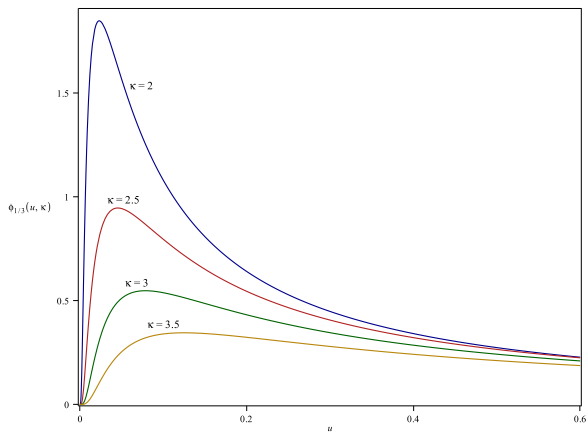}
\caption{\label{fig2} (Color online) Comparison of $\phi_{1/3}(u, \kappa)$ for $\kappa~=~2, 2.5, 3$, and $3.5$.}
\end{center}
\end{figure}

Eq. \eqref{E16} with the ``initial'' condition $\varPhi_{\alpha}(u, 0) = f(u)$ is the Cauchy problem, whose formal solution for $u > 0$ and $0 < \alpha < 1$ can be represented as
\begin{equation}\label{E17}
\varPhi_{\alpha}(u, \kappa) = \int_{0}^{\infty} \phi_{\alpha}(\xi, \kappa) f(u-\xi) d\xi,
\end{equation}
as shown in \cite{KGorska12}. The kernel $\phi_{\alpha}$ has a scaling form
\begin{equation}\label{E18}
\phi_{\alpha}(u, \kappa) =\kappa^{-1/\alpha}\varPhi_{\alpha}(u \kappa^{-1/\alpha}, 1)
\end{equation}
and is the solution of Eq. \eqref{E16} for $f(u) = \delta(u)$, the Dirac delta, see Fig. \ref{fig2}. The $u\to\infty$ asymptotics of \eqref{E18} is given by $u^{-(1+\alpha)}$ which agrees with the experimental data \cite{GZatryb11}. 

The alternative way of expressing Eq. \eqref{E17} is using the evolution-like operator \cite{KGorska12}
\begin{equation}\label{E19}
\hat{U}_{\alpha}(\kappa) = \exp(- \kappa \partial^{\alpha}_{u}) = \int_{0}^{\infty} \phi_{\alpha}(\xi, \kappa) e^{- \xi \partial_{u}} d\xi
\end{equation}
whose action on $f(u)$ gives $\varPhi_{\alpha}(u, \kappa)$. From Eq. \eqref{E9} it is easy to show that for $0 < \kappa_{1} < \kappa_{2}$, we have
\begin{equation}\label{E20} 
\hat{U}_{\alpha}(\kappa_{2}) \hat{U}_{\alpha}(\kappa_{1}) = \hat{U}_{\alpha}(\kappa_{1} + \kappa_{2}).
\end{equation}
That substantiates the semigroup property of KWW law, \textit{i.e.}
\begin{equation}\label{E4}
\frac{n(t)}{n_{0}} = \frac{n(t)}{n(t_{1})} \frac{n(t_{1})}{n_{0}}, \quad 0 < t_{1} < t.
\end{equation}

\section{Discussion and Conclusions}

In this paper we have provided at least two means of analysis of PL decay phenomena. First, mainly of theoretical nature, makes use of exact forms of one-sided L\'{e}vy laws. It appears to be the first application of these laws for non-trivial rational values of the exponent $\alpha$: in this study $\alpha = 17/25$ and $\alpha =11/20$ gave very satisfactory agreement with the experiments. Thus it offers the possibility of analyzing the experimental data avoiding methods based on numerical algorithms for Laplace transforms. The second is an educated guess that the process in question is ruled by a sub-diffusive dynamics mathematically expressed by fractional partial differential evolution equations. These are known to govern other stochastic fractal processes \cite{JKlafter11}. This aspect has also been discussed in \cite{WTCoffey06} where it has been shown that the scaling properties we have dealt with are manifestation of such a behavior. 

Let us observe that Eq. \eqref{E18} has exactly the same form as Eq. (35) on p. 33 in \cite{WTCoffey06}. The scaling behavior of $\phi_{\alpha}(u, \kappa)$ can be confirmed by the diffusion entropy analysis based on the Shannon entropy $S(t)$. In our case, the Shannon entropy calculated for $\phi_{\alpha}(u, \kappa)$ has the identical nonlinear form as Eq. (91) on p. 47 in \cite{WTCoffey06}, namely
\begin{equation}\label{E21}
S(\kappa) = - B + \frac{1}{\alpha} \ln(\kappa),
\end{equation}
where the constant $B$ is determined by the time-independent LSD distribution 
\begin{equation}\label{E22}
B = - \int_{-\infty}^{\infty} \varPhi_{\alpha}(y; 1) \ln \varPhi_{\alpha}(y; 1) dy.
\end{equation}
That gives the fractal scaling index equals to $1/\alpha$ \cite{NScafetta02}. Furthermore, in multidimensional disordered systems it is shown that the relaxation processes are of fractal nature \cite{WTCoffey06, ZCheng87} and are related to the percolation \cite{WTCoffey06, AVMilovanov01, AVMilovanov02}. All of these facts enable us to conjecture that processes underlying PL could indeed be of a fractal nature. We would like to remark that from mathematical point of view these curious properties of the sample are strictly connected with the stretched exponential behavior of PL.

Any further study aimed at providing a significant progress in this field should consider a disordered medium (nano-crystals) of a mesoscale size in non-equilibrium state. To correctly describe the behavior of our medium we should use the combined method of quantum and non-equilibrium statistical mechanics. At present, such methods are not available. From our point of view such a necessary mathematical tool should emerge from the the deeper understanding of physics of fractional calculus and its relation to LSD. According to our discussion most of the puzzling features of the PL decay phenomenology might be traced back to the sub-diffusive nature of L\'{e}vy flights in Si-NCs. It confirms that the explicit use of LSD for general rational $\alpha$ is a powerful tool in interpretation of experimental data.

\section*{Acknowledgments}

We thank Profs. M. Massalska-Arod\'{z}, J.~Adamowski, G. Baldacchini, Dr. Francesca Bonfigli, and M. Perzanowski  for important discussions.

We gratefully acknowledge the anonymous referee for constructive advices and suggestions which have contributed to improve this paper.

The authors acknowledge support from the PHC Polonium, Campus France, project no. 28837QA.



\begin{thebibliography}{20}
\bibitem{CAS} Here is Maple$^{\text{\textregistered}}$ procedure quoted from \cite{KAPenson10} {\tt LSD(k, l, x)} used to calculate $\varPhi_{l/k}(u)$:
{\tt b~:=~proc(j,k,l) simplify(product(GAMMA((i-j)/k),i=1.. j-1) * product(GAMMA((i-j)/k),i=j+1..k-1) * (l$\land$(l/k)/k)$\land$j / product(GAMMA(i/l-j/k),i=1..l-1)); end;} \\
{\tt LSD:=(k,l,x) sqrt(k*l)/(2*Pi)$\land$((k-l)/2) * sum(b(j,k,l)*x$\land$(-j*l/k) * hypergeom([1, seq(1/l + j/k + i/l, i=0..l-1)],[seq((1+j+r)/k, r=0..k-1)], (-1)$\land$(k-l)*l$\land$l/(k$\land$k*x$\land$l))/x, j = 1..k-1); end;
}
\end{thebibliography}
\end{document}